\documentclass[sigconf,nonacm]{acmart}
\usepackage{subfigure}
\usepackage{multirow}
\usepackage{booktabs}
\usepackage{subcaption}

\AtBeginDocument{%
  }

\begin{document}

\title{CompassDB: Pioneering High-Performance Key-Value Store with Perfect Hash}


\author{Jin Jiang}
\authornote{Corresponding author}
\authornote{Both authors contributed equally to the paper}
 \email{jin@dipeak.com}

\affiliation{%
  \institution{DiPeak}
   \city{Beijing}
   \country{China}
}

\author{Dongsheng He}
\orcid{0009-0009-4729-3078}
\authornotemark[2]
\affiliation{%
  \institution{DiPeak}
  \city{Chengdu}
   \state{Sichuan}
   \country{China}
}

\author{Yu Hu}
\authornotemark[2]
\affiliation{%
  \institution{DiPeak}
   \city{Chengdu}
   \state{Sichuan}
   \country{China}
}

\author{Dong Liu}
\affiliation{%
  \institution{DiPeak}
  \city{Wuhan}
  \state{Hubei}
  \country{China}
}

\author{Chenfan Xiao}
\affiliation{%
  \institution{DiPeak}
   \city{Chengdu}
   \state{Sichuan}
   \country{China}
}

\author{Hongxiao Bi}
\affiliation{%
  \institution{DiPeak}
   \city{Chengdu}
   \state{Sichuan}
   \country{China}
}

\author{Yusong Zhang}
\affiliation{%
  \institution{DiPeak}
   \city{Chengdu}
   \state{Sichuan}
   \country{China}
}

\author{Chaoqu Jiang}
\affiliation{%
  \institution{DiPeak}
   \city{Chengdu}
   \state{Sichuan}
   \country{China}
}

\author{Zhijun Fu}
\affiliation{%
  \institution{DiPeak}
   \city{Beijing}
   \country{China}
}

\renewcommand{\shortauthors}{Trovato et al.}

\begin{abstract}
Modern mainstream persistent key-value storage engines utilize Log-Structured Merge tree (LSM-tree) based designs, optimizing read/write performance by leveraging sequential disk I/O. However, the advent of SSDs, with their significant improvements in bandwidth and IOPS, shifts the bottleneck from I/O to CPU \cite{lepers2019kvell}. The high compaction cost and large read/write amplification associated with LSM trees have become critical bottlenecks. In this paper, we introduce CompassDB, which utilizes a Two-tier Perfect Hash Table (TPH) design to significantly decrease read/write amplification and compaction costs. CompassDB utilizes a perfect hash algorithm for its in-memory index, resulting in an average index cost of about 6 bytes per key-value pair. This compact index reduces the lookup time complexity from $O(log N)$ to $O(1)$ and decreases the overall cost. Consequently, it allows for the storage of more key-value pairs for reads or provides additional memory for the memtable for writes. This results in substantial improvements in both throughput and latency. Our evaluation using the YCSB benchmark tool shows that CompassDB increases throughput by 2.5x to 4x compared to RocksDB, and by 5x to 17x compared to PebblesDB across six typical workloads. Additionally, CompassDB significantly reduces average and 99th percentile read/write latency, achieving a 50\% to 85\% reduction in comparison to RocksDB. 
\end{abstract}

\maketitle

\section{Introduction}

Persistent key-value storage plays a vital role as a key part of data storage infrastructure. Firstly, the Key-value interface is so generic that applications can have more flexibility to store their data in various schemas. Secondly key-value pairs are a simple data model that can provide high-performance read and write operations. Thirdly, key-value is easy to horizontally scale. These advantages of key-value storage make it popular for a wide range of use cases, such as a storage engine for databases (e.g., CockroachDB \cite{taft_cockroachdb_2020}, MyRocks\cite{matsunobu_myrocks_2020}, MongoDB \cite{mongodb_nodate} ),  a logging/queuing service (e.g., RocketMQ \cite{apacherocketmq_nodate} ),  a caching storage on SSD (e.g., speedb \cite{speedb-iospeedb_2024}, kvrocks \cite{apachekvrocks_nodate}).  

For data-intensive services, the performance of key-value storage is crucial to response latency and throughput. Numerous research efforts have been dedicated to speed up key-value stores performance. RocksDB \cite{dong2017optimizing, dong_evolution_nodate} is derived from LevelDB \cite{googleleveldb_2024} and optimized for fast, low latency storage such as flash drives and high-speed disk drives.  PebblesDB \cite{raju_pebblesdb_2017} uses a fragmented log-structured merge trees to reduce write amplification. SplinterDB \cite{conway2020splinterdb,conway2023splinterdb}, based on  b-epsilon tree, maximizes bandwidth utilization of modern storage devices.  TreeLine \cite{yu_treeline_2022}, an update-in-place kv store using in-memory tree index and insert forecasting technique.

From our practical experience regarding key-value storage, the performance of point lookup is often more crucial than scans in most scenarios. Simultaneously, the read-write amplification introduced by LSM-Tree is a significant factor affecting read and write performance \cite{254461,strategies_nodate}. 


Based these observations, in this paper, we proposed a novel data structure, Two-tier Perfect Hash Table (TPH) , which combine Perfect Hash \cite{czech1997perfect} and Log-Structured Merge Trees (LSM-Tree) \cite{o1996log}. TPH can achieves constant lookup time cost ($O(1)$) benefit from characteristics of perfect hash. Meanwhile TPH using piece mechanism to reduce write amplification further to significantly improve read/write efficiency.

A TPH table consists of a group of piece files, each storing a range of key-values, and can be regarded as equivalent to an SSTable in LevelDB. When new data arrives, a new piece file is generated to store the changed data, while the unchanged data remains in the old file. Each piece file is read-only once created. This mechanism significantly reduces write amplification compared to the traditional compaction method in LSM-Trees.

Additionally, TPH uses a perfect hash algorithm to generate two indexes. The global index maps keys to piece files, and the local index locates the offset within the file where the key resides. The index information is written into the metadata of the newly added piece file, facilitating subsequent data searches. TPH optimizes space usage for the perfect hash function, allowing all metadata to be retained in memory. As a result, each query requires only two index lookups in memory and one disk I/O to access the data, achieving excellent read performance with a time complexity of $O(1)$, which is lower than the $O(log n)$ of B-trees and SSTables. More details will be discussed in Section \ref{TPH}.

Built on top of RocksDB, an industrial-grade key-value storage engine with high throughput and low latency, CompassDB utilizes the TPH data structure. Like LSM-Trees, CompassDB organizes data into multiple levels with increasingly larger, sorted components (TPH) beyond level 0. CompassDB also introduces the "hash range" model, where each level is evenly partitioned into a fixed number of components based on the hash of the key, and the number of components in each layer increases exponentially. This model splits the entire tree into multiple independent sub-trees, designed for fast lookups and reduced write amplification.

CompassDB is compatible with all RocksDB APIs, allowing users to easily migrate from existing RocksDB applications with minimal changes. It also supports various data schemas, such as wide-table maps and graphs, to accommodate more complex business scenarios.

In our comparisons, CompassDB outperforms RocksDB under most load conditions. The write amplification of RocksDB is about twice that of CompassDB, with the ratio increasing as the dataset grows. Space amplification is 1.2 times that of CompassDB. In a pure read scenario, RocksDB's read amplification is three times that of CompassDB. The average and 99th percentile latencies of CompassDB are one-half to one-third of RocksDB, while CompassDB's throughput is two to three times higher than that of RocksDB.

In summary, this paper makes the following contributions:

\begin{itemize}
    \item  A much more efficient method for computing perfect hash.
    \item  The novel design of TPH data structure, offering stable read and write performance while significantly reducing the latency and increase the throughout. 
    \item Designs and implements CompassDB, an industrial-grade high-throughput KV engine that has been successfully applied in services for millions of users.
\end{itemize}
\section{Background}

This section introduces the basic background of Log-Structured Merge Tree (LSM-Tree) and addresses the issues of write and read amplification that arise from its use. Additionally, it provides a brief introduction to RocksDB, a database built on the LSM-Tree, and an explanation of perfect hash algorithms.

In the field of persistent key-value storage, the LSM-Tree, on the other hand, is optimized for write-heavy workloads. When data update operations (writes, deletions, or updates) occur, they are first placed into an in-memory memtable, allowing for faster writes compared to disk. Once the memtable reaches a certain size, it triggers a background flush job, writing data to the disk in the form of an SSTable at level 0. Periodically, a merging and compaction process runs in the background, combining SSTable files between consecutive levels into new SSTables. The SSTables above level 0 are strictly ordered. To serve a read request, the system first checks the memtable, then performs a linear search in the most recent SSTable at level 0, followed by a binary search in subsequent levels. Additional filters, such as Bloom filters \cite{bloom1970space}, are often used to determine whether a key exists in an SSTable, avoiding unnecessary reads.

The LSM-Tree generally outperforms the B+ Tree \cite{comer1979ubiquitous, wiredtiger_nodate, olson1999berkeley} in terms of write performance by increasing write throughput through sequential writes and background merge operations. However, this design introduces issues of read amplification and write amplification. Read operations may need to access data across multiple levels, and write operations may require data to be written to disk multiple times.

As data is continuously written to the LSM-Tree, it becomes tiered, with data progressively sinking down through the levels. This tiered mechanism optimizes write performance but inevitably leads to write amplification. Specifically, whenever data is merged from one tier to the next, all overlapping data needs to be rewritten to disk, even if the data has not changed. This process significantly increases the amount of data written to disk compared to the actual data size.

As data is continuously written to the LSM-Tree, it becomes tiered and data is progressively sunk down. 
This tiered mechanism is helpful for optimizing write performance, but inevitably leads to the phenomenon of write amplification. Specifically, whenever data is merged from one tier to the next, all overlapping data needs to be rewritten to the disk, even if the data has not changed. This process significantly inflates the amount of data written to the disk compared to the actual data size.

For storage devices with limited write lifespan, such as Solid State Drives (SSDs), frequent write operations can degrade their lifespan more quickly \cite{agrawal2008design,grupp2009characterizing,mielke2008bit,narayanan2009migrating}, 
, increasing maintenance costs.

Read amplification refers to the ratio of the total amount of data read from storage to the actual amount of data requested by the user. The read process in an LSM-Tree starts from the memtable and searches down through each level until the target is found. This process may involve multiple disk accesses, as each tier may have an SSTable that needs to be checked. Each access to an SSTable requires reading its metadata, such as index blocks and filters. False positives from filters can also lead to unnecessary data block reads, increasing the amount of data accessed, especially in scenarios with large volumes of data and complex access patterns.

\subsection{RocksDB}

RocksDB is a high-performance key-value storage engine inherited from LevelDB, with support for persistent data storage, optimizations for concurrent writes and multi-thread access. RocksDB uses the LSM-Tree as its data storage structure, effectively reducing the random I/O demands on the disk from write operations. The architecture of RocksDB allows it to recover data after a system crash and supports high-throughput data processing, while offering data compression options and optimization strategies to further improve storage efficiency and
read performance. However, RocksDB also faces the inherent issues of read amplification and write amplification associated with the LSM-Tree. Read amplification occurs during read operations, which require accessing data files across multiple tiers, with the file access efficiency being $O(LOG(N))$, further reducing read efficiency. Write amplification occurs during the background compaction process, where unmodified data is rewritten multiple times, increasing the amount of written data and slowing down the compaction speed. Therefore, compaction operations can cause fluctuations in RocksDB’s performance. Despite this, RocksDB is cross-platform capabilities and support for multiple programming languages enable developers to integrate it into various applications. The ecosystem built around RocksDB also provides a wealth of tools and services, including cloud database services and third-party libraries, greatly expanding the application scope and flexibility of RocksDB. Based on these advantages, we have chosen to build CompassDB on top of RocksDB, ensuring full compatibility with RocksDB’s interface.

\subsection{Perfect Hash}

Perfect hash is an efficient mapping function that can map a definite set of keys to unique integer indices without any hash collisions. The Minimal Perfect Hash (MPH) can map n keys precisely to a continuous integer interval of size n, thus achieving optimal space utilization. This is achieved by maintaining a load factor of 100\%, which is the ratio of the number of elements stored within the hash table to its total capacity.

CHD \cite{belazzougui2009hash} and PTHash \cite{pibiri2021pthash} are two common perfect hash algorithms. Their construction processes are similar, both carried out in three stages namely mapping , ordering and searching.

\begin{itemize}
    \item \textbf{Mapping} Mapping keys into m buckets.
    \item \textbf{Ordering}. Sorting the buckets by non-increasing size to speed up the searching step.
    \item \textbf{Searching}. For each bucket $m_i$ in the order given by the b) step to find a set of parameters P that make all keys occupied positions not conflict with previous keys. 
\end{itemize}

For the CHD algorithm, keys are uniformly dispersed among the buckets in the mapping stage .
In the searching stage,  CHD algorithm uses two generic hash functions to calculate two hash values for each key: $h_0$, $h_1$, where $n$ represents the bucket number, and $h_0$, $h_1$ are used to calculate the offset position in the perfect hash array. The mapped position is:

\begin{equation}
    \label{ph_position_equation}
        position = (h0_i + (h1_i \times \alpha_i) + \beta_i) \mod tablesize
\end{equation}

Here, $\alpha_i$ and $\beta_i$ represent the parameters in the $i-th$ bucket. $h0_i$ and $h1_i$ represent the hash value for the two hash functions. The algorithm continuously tries search  and for each bucket to ensure that all positions are conflict-free.

CHD and PTHASH are two relatively common perfect hashing algorithms. Their construction processes are similar: first, by hashing n keys into m buckets; then, conducting an exhaustive search for suitable hash function parameters for each bucket to ensure that the keys do not conflict after hashing; finally, compressing and encoding the parameters for each bucket and storing them in an array P, which represents the perfect hash of the key set.

Different from the CHD in the mapping stage, the PTHash algorithm divides the buckets into dense buckets $d$ ($0.3m$) and sparse buckets $S2(m - d$)

Mapping  60\% of the keys ($S1$) to dense buckets, and the remaining 40\% ($S2$) of the keys to the sparse buckets. The mapping algorithm formula is as follows:

\begin{equation}
    \label{equ:mapping}
    \text{bucket}(key) = \begin{cases}
        h(key, s) \bmod d, & \text{if } key \in S1 \\
        d + h(key, s) \bmod (m - d), & \text{otherwise}
    \end{cases}
\end{equation}

Benchmarks have revealed that PTHash has an advantage over CHD in terms of performance of construction, especially when dealing with large amounts of data. CHD algorithm is better space efficiency compared to PTHash. We design CPHash take advantage of CHD and PTHash to balance space and construction time efficiency.(see Section \ref{CPHash})
\section{CompassDB Design}

In this section, we will introduce the core design of CompassDB, which aims to reduce write and read amplification while maintaining stable and excellent read performance. To achieve this, CompassDB's design includes two main elements. First, it features an index structure based on a perfect hash algorithm, optimized for spatial and construction time efficiency, providing an access time complexity of $O(1)$. Second, it uses a mechanism involving piece files to minimize unnecessary data rewrites, thereby reducing write amplification.

\subsection{CPHash}
\label{CPHash}

Leveraging the strengths of CHD and PTHash, we have implemented CPHash, a perfect hash construction algorithm. CPHash categorizes CHD’s buckets into dense and sparse buckets. To minimize dictionary entries, it attempts to reuse parameters from existing buckets when constructing new ones, reducing the number of searches and saving storage space.

We use DJB \cite{djb_hash} hash and city hash \cite{googlecityhash_nodate} as the default two hash functions(termed $h1$ , $h2$). DJBHash is simple and fast and CityHash provides high hash quality.

\begin{figure}[t]
  \centering
  \includegraphics[width=\linewidth]{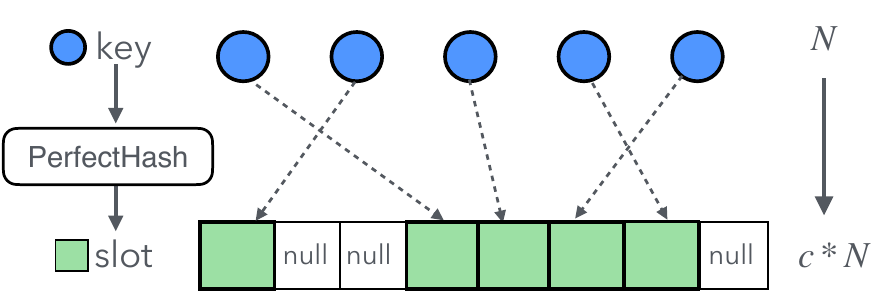}
  \caption{The space of $N$ keys is mapped to an array of slots with a length of $N*c$, where each key is perfectly hash to a unique slot. The length of the slot array is greater than the number of keys.}
  \Description{}
  \label{fig:CPHash_mapping}
\end{figure}

So, in the mapping stage we mapping 60\% entries to dense buckets and remaining to sparse buckets, and using CityHash as the hash function at equation \ref{equ:mapping}. And we scale the $tablesize$ by $c$ at equation \ref{ph_position_equation} in searching stage. Additionally, in the implementation, we use SIMD instructions to enhance  computational efficiency. 

A MPH function can map $N$ distinct keys to the range [1, $N$] without having any collisions, which achieve optimal space utilization. In practice, attempting to achieve a 100\% load factor will consumes more computational resources and increases the searching time. Therefore, CompassDB using CPHash mapping $N$ keys to the range [1, c * $N$] as shown in 1, where $c$ is the scale factor, with a default value of 1.1 (load factor = 90\%), trading off space for time efficiency.

\subsection{Two-tier Perfect Hash Table}
\label{TPH}

In LSM-tree, write amplification is primarily attributed to the redundant writing of data. Two-tier Perfect Hash Table (TPH) is designed stems from the intuition that data remains unmodified should not be migrated, the altered data is all you need to write.

One TPH is a group consisting of multiple piece files, which logically corresponds to an SSTable file in RocksDB. Each piece contains a portion of the data from the table. When new data comes in (compaction), a new piece file will be created to store these updated data, and replace the previous most recent piece as the head piece. At the same time, the head piece also acts as the index holder for all keys, indicating in which piece the up-to-date version of each key resides. The traditional hash table is inefficient for persistent storage for several reasons: 1) In-Place Updates consume numerous random disk I/O operations; 2) When the hash table's capacity is insufficient, resizing requires reassigning data to new positions; 3) Resolving conflicts necessitates multiple I/O attempts; 4) Hash tables do not support range queries. Consequently, most storage engines for persistent storage are now based on B-tree series or LSM-tree structures, but the point lookup time complexity for both is $O(Log(N))$.

The LSM-tree framework is particularly well-suited for use with hash tables. Firstly, LSM-tree write data sequentially, avoiding random disk I/O. Secondly, all data in the SSTables of LSM-trees is read-only once generated, eliminating the need for rehashing. Thirdly, perfect hash algorithms can be used to make hash tables conflict-free, resolving issues related to conflicts.

Based on these principles, we can use a perfect hash table to replace the SSTable of an LSM-tree as the basic storage unit on SSDs. During compaction or flushing, the selected key set is fixed, allowing the construction of a perfect hash table for this set and assigning key-values to corresponding positions in the file. Compared to RocksDB SSTables, perfect hash table indexes are small (usually several bits per key) and proportional to the number of keys, whereas SSTable index sizes are proportional to the data size. Thus, the perfect hash (PH) index can fit into memory, speeding up index lookups(see Section \ref{ph_index_in_memory}.

In LSM-trees, write amplification primarily arises from redundant data writing. The Two-tier Perfect Hash Table (TPH) is designed based on the intuition that unmodified data should not be migrated; only altered data needs to be written. A TPH consists of multiple piece files, which logically correspond to an SSTable file in RocksDB. Each piece contains a portion of the data from the table. When new data is added (during compaction), a new piece file is created to store these updates, replacing the previous most recent piece as the head piece. The head piece also acts as the index holder for all keys, indicating the piece where the up-to-date version of each key resides

\begin{figure}[t]
  \centering
  \includegraphics[width=\linewidth]{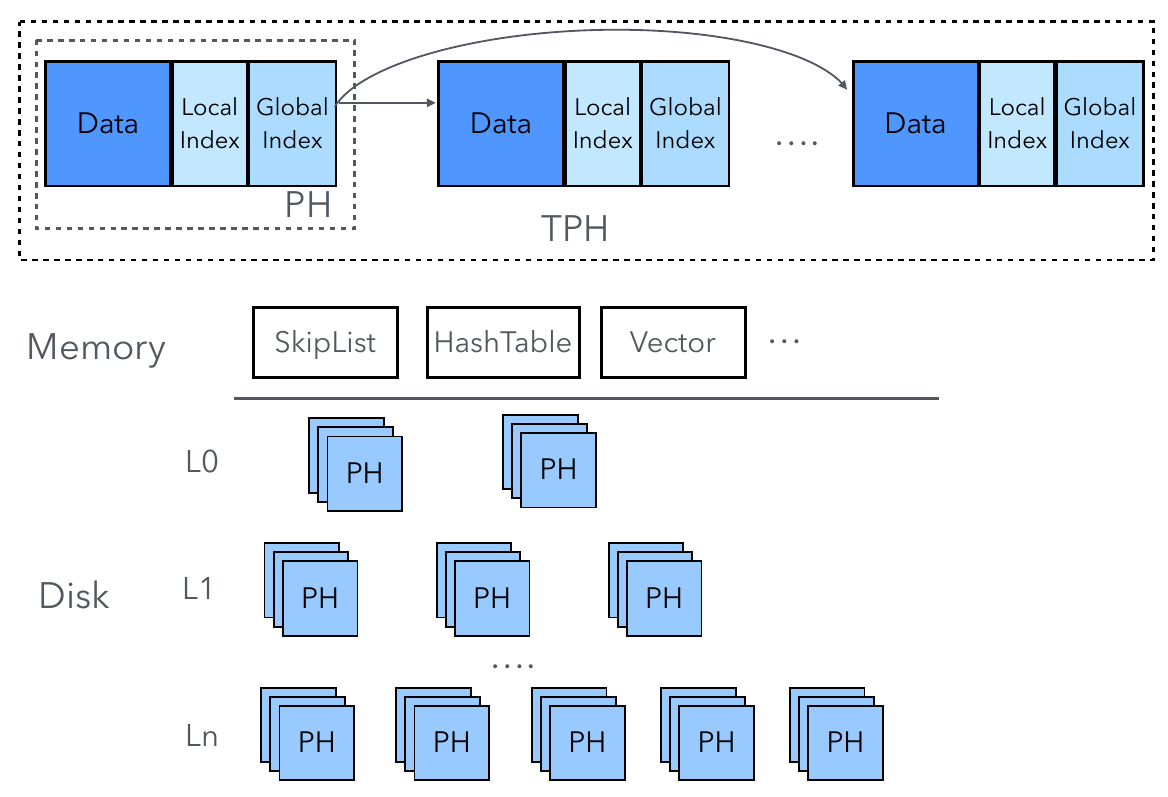}
  \caption{Each TPH is equivalent to the position of SST in RocksDB, with each TPH composed of multiple piece files, each piece file containing its own local hash table; TPH includes a global hash table pointing to the piece file where the actual Key is located.}
  \Description{A woman and a girl in white dresses sit in an open car.}
    \label{fig:img-tph}
\end{figure}


\subsection{Operations}
\begin{figure*}[t]
  \centering
  \includegraphics[width=0.8\linewidth]{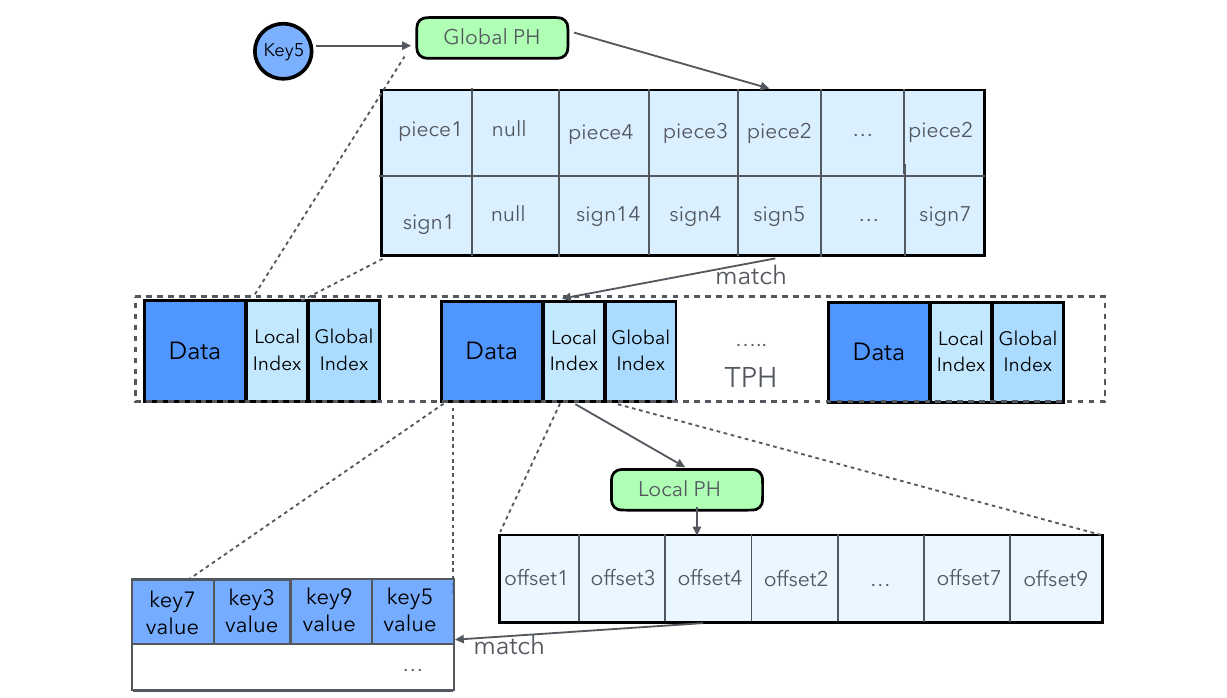}
  \caption{K5 is mapped to the position with index=4 in the global hash table (index starts from 0), and signature matches sign5, the search process continues. The piece number where K5 is located is identified as piece file 2. Within piece 2's local hash table, the slot position corresponding to K5 is recalculated based on the local hash table, it yields offset-4. Then we read from the piece file 2 at the position offset-4 to get the slot value. If key of slot is matched, we got the final value.}
  \Description{A woman and a girl in white dresses sit in an open car.}
  \label{fig:lookup}
  
\end{figure*}

\textbf{Lookup}: Lookup: As shown in Figure \ref{fig:lookup}, TPH is divided into two perfect hash tables: the global perfect hash table within the piece group and the local hash table within each piece file. The global hash table uniquely maps the 64-bit hash value of a key to a slot in a one-dimensional array. This slot stores two pieces of information: the key's signature and the piece file where the key resides. The key signature is similar to a Bloom filter. When the signature in a slot matches the signature of the key being searched, it highly likely indicates the presence of the key in the TPH. For each key in the dataset, a hash function computes a \verb|uint8| value as the key’s signature. When loading historical data from the database files, this signature array is loaded into memory. During a key lookup, the perfect hash function first calculates the position of a key in the global hash table and its corresponding signature. This signature is then compared with the signature at the corresponding slot. If the signatures match, the piece file containing the key is identified, allowing for efficient retrieval of the key position in the signature array. When the signatures do not match, it indicates that the key definitely does not exist. When the signatures match, the key likely exists, with a false positive probability of 1/255, meaning within  $8 * 1.15 = 9.2$  bits of memory, the false positive rate is 0.392\%. In comparison, RocksDB defaults to using a 10-bit bloom filter, with a false positive probability of 0.812\%. 

Another information is the pointer to the target piece file. The local hash table within the piece file records the actual location of the KV data in target piece file. Within the piece file, we calculate the slot index of the key in the local hash table to retrieve the storage position in the piece. Then read the key-value pair and compare it with the searched key to see if it is consistent.
Given the compact memory footprint of the perfect hash tables, both the global and local hash tables of all TPH instances can reside entirely in memory. This enables the entire indexing process to be conducted within memory, requiring only a single disk I/O operation to read the key-value pair. In contrast, managing large datasets in RocksDB, which may span numerous SST files, poses challenges because its index data cannot be fully accommodated in memory. As a result, accessing index and data in RocksDB often necessitates multiple disk I/O operations. Experimental findings indicate that RocksDB typically incurs 2-3 times more disk IOPS and handles larger disk I/O sizes per access compared to CompassDB.

\textbf{Scan}: In CompassDB, leveraging perfect hashing results in a random distribution of data within each TPH (Two-phase Hashing). To support iterating through keys within a specified range, CompassDB employs a sorting strategy during the final stage of data compression. All keys within the TPH are sorted based on a user-defined comparator (typically alphabetical order by default). Post sorting, index information corresponding to each key is recorded and written into the piece file. This approach involves marking each key with a 64-bit integer for its slot information, effectively reversing the order of the hash index to align with user-specified data ordering.

\begin{figure}[t]
  \centering
  \includegraphics[width=\linewidth]{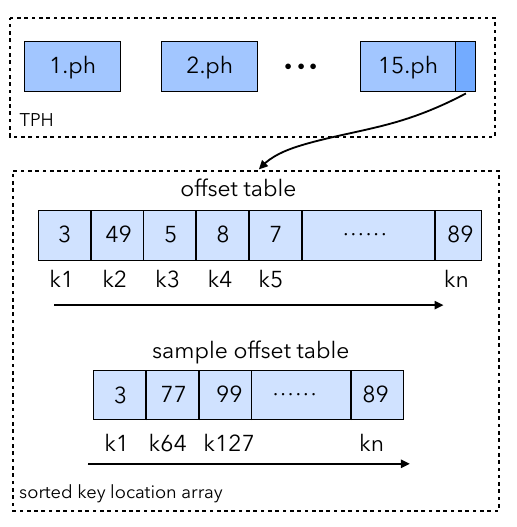}
  \caption{The sorted key location array. The latest piece file record the offset of sorted keys within current TPH. And use interval sample to improve scan performance.}
  \Description{}
  \label{fig:scan}
\end{figure}

Figure \ref{fig:scan} illustrates that the most recent piece file in each TPH contains a reverse index of all keys within that TPH. This reverse index details the slot position information of keys sorted in the global hash table, facilitating efficient range queries. Furthermore, to optimize scan operation performance, CompassDB samples a key every N keys (default interval is 64 keys) from the index information array and stores these samples in the piece file. During a scan operation, the system initially loads these sampled key arrays from disk into memory. This preloading aids in swiftly locating the starting key for the scan, thereby enhancing data retrieval efficiency.

\textbf{Delete}: Deleting an existing key in CompassDB involves marking it with a placeholder to signify its deletion. This placeholder remains in place and affects global hash calculations until the associated piece file of the old key undergoes garbage collection. Only after this process is completed can the key be permanently removed, effectively managing the issue of handling deleted keys.
With this design, even there is a signature conflict, the system can correctly identify which key-value pairs have been deleted, thus avoiding the problem of misreading deleted data. This method not only ensures data consistency but also maintains the integrity of the perfect hash index.

\subsection{Hash Range}

CompassDB introduces an innovative data indexing mode known as Hash Range, designed specifically for environments with frequent write operations and high-performance point queries. The fundamental concept of Hash Range involves hashing the original keys to generate a 32-bit integer termed the Search Key. In this mode, the distribution of key-value pairs is determined by this Search Key rather than the original key.

\begin{figure}[t]
  \centering
  \includegraphics[width=\linewidth]{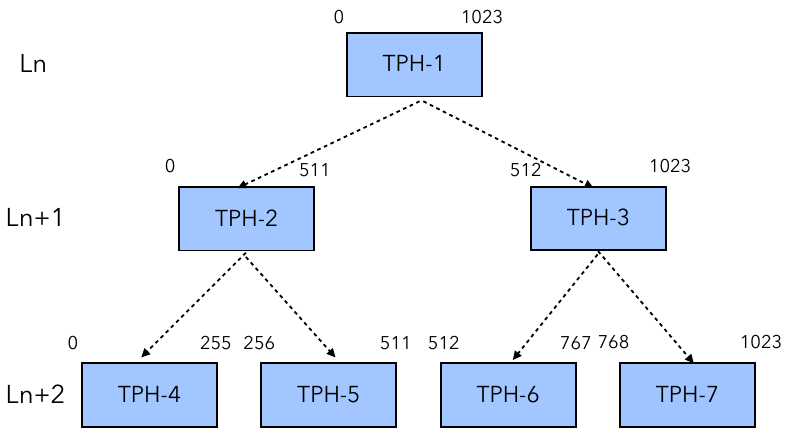}
  \caption{The range of keys contained in TPH-1 is the search key in the range [0, 1024). Only overlap with 2 TPHs in the next level.}
  \Description{Figure 8: }  
  \label{fig:hash range}
\end{figure}

Figure \ref{fig:hash range} illustrates that each TPH in Hash Range no longer manages the range of original keys but instead operates within the range of Search Keys. This organizational approach forms a hierarchical structure resembling a tree. During compaction operations, each TPH at every level, except Level-0, divides its data within the Search Key range into N partitions. These partitions are then merged with corresponding TPHs in the subsequent layer. For instance, if the full Search Key space spans from 0 to 1023, TPH-1 might contain key-value pairs with Search Keys in the range [0, 1024). During compaction, TPH-1 could overlap with at most two TPHs in the next level, such as TPH-2 [0, 512) and TPH-4 [512, 1024). After compaction, TPH-1's data is split into two delta piece files, which are then merged into TPH-2 and TPH-3 respectively, according to their Search Key ranges. This structured approach ensures a fixed number of TPHs participate in each compaction operation.

Hash Range mode optimizes compaction processes by precisely controlling the involvement of files, thereby minimizing unnecessary data rewriting and reducing read and write amplification. It achieves this by isolating different "subtrees," enhancing concurrency during compaction and thereby improving overall write performance significantly. The effectiveness of Hash Range mode is analogous to the guard mechanism in PebbleDB. To validate its benefits, tests were conducted using a dataset of 400 million keys with a value size of 1KB, detailed in Table 2. Results showed a 57\% reduction in data read during compaction, a 49\% decrease in compaction time, and an overall 55\% reduction in running time.

\subsection{Compaction}
\begin{figure*}[t]
  \centering
  \includegraphics[width=0.8\linewidth]{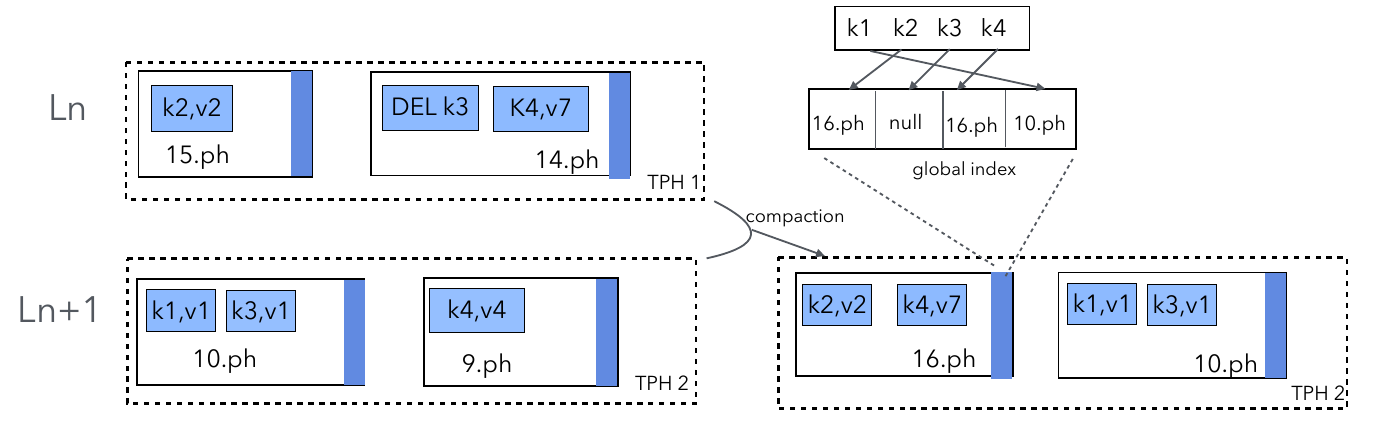}
  \caption{Compaction process. k2 is the newly added key, it will be asigned to the newly piece file(16.ph). k1 is not changed, it still lies in 10.ph. k3 is deleted, but 10.ph is still live, it will be recorded in global index but point to invalid piece. k4 is updated, it will be move to the newly piece file, and 9.ph has no valid data, it would be released.}
  
  \Description{Figure 8: The range of keys contained in TPH-4 is the search key in the range [255, 512).}
  \label{fig:img-7}
\end{figure*}

In CompassDB, compaction operations are measured in terms of TPH units both for input and output. Figure \ref{fig:img-7} illustrates a compaction process under the hash range mode. Here, TPH-1 from Level 1 serves as the delta TPH, while TPH-2 from the next level acts as the base TPH. CompassDB utilizes these TPHs to gather all valid key-value pairs, constructing a new piece file within the base TPH.

Initially, CompassDB creates an array in memory called MergeArray. This array facilitates the insertion of keys from both the delta TPH and the base TPH into specific slots based on their index values derived from the perfect hash table. The slots in MergeArray can assume one of four cases:

\begin{enumerate}
    \item Both delta and base are empty.
    \item Only delta has values.
    \item delta and base have values.
    \item Only base has values.
\end{enumerate}

For cases (2) and (3), where delta and base have values, these are merged, and the updated key-value pairs are stored in the new piece file. For example, in Figure 6, key k2 is directed to the newly generated piece file by the global hash table. In case 4, where only base data exists and remains unchanged, it remains in its original piece file without modification (e.g., key k1 in Figure\ref{fig:img-7} remains in file 10.ph).

Once construction is complete, CompassDB generates a new global hash table based on MergeArray. For non-empty slots in MergeArray, the global hash table replicates the key signatures and records the piece file index where each key resides. As illustrated, piece file 16.ph encompasses the entirety of the write-in volume for this compaction, containing only the key-value pairs requiring updates, while files like 10.ph remain unchanged.

However, reducing write amplification inevitably leads to space amplification as TPH undergoes multiple compactions. Over time, old piece files may accumulate invalidated key-value pairs. To manage this, CompassDB limits the number of piece files per TPH (default limit is 16, configurable up to 1024). When this threshold is exceeded, the system triggers garbage collection. During garbage collection, the oldest piece files and those with a significant proportion of invalidated key-value pairs are marked for update. In the subsequent compaction, all valid key-value pairs from these files, along with any newly added, updated, or deleted pairs from the previous level, are consolidated into a newly generated delta piece file. Once compaction is complete, these old files can be safely deleted. For example, in Figure 6, when key k3 is deleted, 9.ph no longer contains valid key-value pairs, prompting CompassDB to automatically remove these files. This mechanism ensures that space amplification remains manageable.

On the flip side, the presence of invalidated key-value pairs increases the amount of data read during compaction operations. The system sequentially reads all piece files stored in TPH to handle these invalidated pairs. However, considering the performance characteristics of SSDs, which excel in read operations compared to writes, leveraging read operations during compaction optimally balances this trade-off. This approach capitalizes on SSD advantages, enhancing overall processing speed during compaction operations.

\section{Implementation}

This section delve into the core implementation of CompassDB includes optimizing CPHash build time, the piece file structure and index memory usage. Additionally, this section will elucidate how CompassDB leverages it’s flexible and diverse configuration options.

\subsection{Optimizing CPHash Build Time}

Section \ref{CPHash} introduce the mathematical algorithm for building CPHash.In fact, the time spent on building a perfect hash table accounts for a proportion of overall time for compaction and flush operations, especially for large scale data.

To expedite the construction of a perfect hash, we also employ the following methods:

    \subsubsection{Additional Space Allocation}  If we can not build the perfect hash table after the limit attempts, to avoid excessive computational overhead, we increase the available hash table slots moderately, map $n$ keys to $ \alpha * n  (\alpha >= 1.0)$ consecutive integers $[0, \alpha * n] $. This approach is based on this assumption that more spaces imply a lower probability of conflicts, thereby decreasing expected number of search retires. 
    
    \subsubsection{Vectorization and Parallelization} CompassDB utilizes vectorization and parallelization techniques to parallelize the calculation of positions. 
    When searching for parameters within a bucket, the position mapping equation \ref{ph_position_equation} is parallelized using Intel Advanced Vector Extensions 2 \cite{intel_nodate} (AVX2 instructions). Because AVX2 do not provide the modular instructions so the modular can not be parallelized directly. To resolve this issue, we rewrote the fastmod \cite {DBLP:journals/corr/abs-1902-01961, lemirefastmod_nodate}, a algorithm that replace an integer division by a multiplication, in AVX2 instructions. Benchmarks shows that this computation process can be accelerated by 2x.

    \subsubsection{Segmentation}  
\begin{figure}[t]
  \centering
  \includegraphics[width=\linewidth]{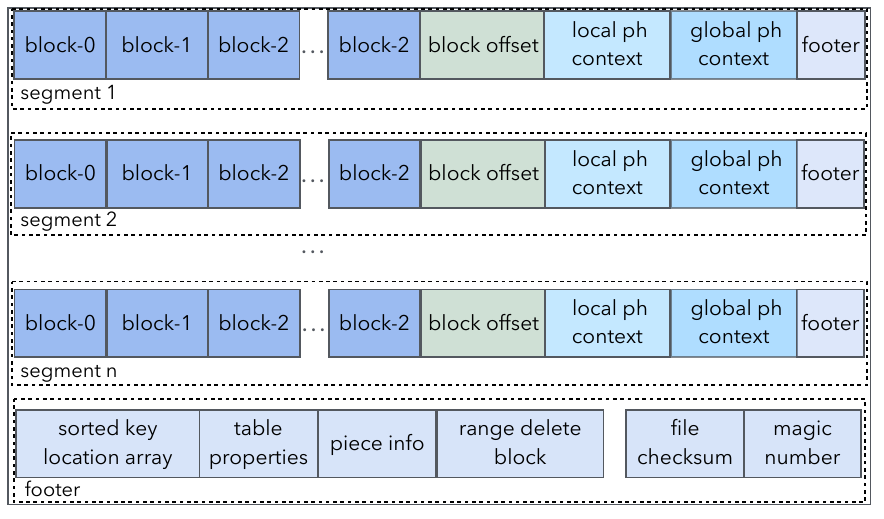}
  \caption{ The layout of on-disk ph file}
  \Description{Figure 8}
  \label{fig:img-ph-structure}
\end{figure}

We observed that the construction time of perfect hash remarkably increases when the number of keys become large. To address this issue, CompassDB divides the piece file  into multiple segments as shown in Figure \ref{fig:img-ph-structure}.
    Each key map into segment by:
    \begin{equation}
        segment(key) = h2(key) \mod m
    \end{equation}
    where $m$ is the segment count (64 by default). Because keys of each segment is independent, so they can do construction job in parallel way to reduce compaction time. RocksDB provides a similar mechanism called \verb|subcompaction| \cite{subcompaction_nodate}, it split keys into multiple parts of approximately equal size by sampling method. Due to the inaccuracy of the sampling, the number of keys distribution in each subset is skewed, which leads to CPU resource are not fully utilized.

Furthermore, to reduce the size of index, we limit the maximum size of a segment to 4GB, 
allowing it to be represented by a 32-bit integer. 
The size of the entire piece file is theoretically unlimited and can be increased by adding more segments as needed.

\subsection{Piece File Structure}

In CompassDB, a TPH serves as a file group comprising multiple piece files, with each piece file having the suffix ".ph". Figure 7 illustrates a typical piece file format, consisting of multiple segments and a file footer. The file footer is responsible for storing metadata information for the entire TPH, including version details, sorting key information, and TPH table attributes. Each segment is responsible for storing user key-value pair data as well as internal indexing information which include the global hash context, piece index table, and signature table.

\begin{figure}[t]
  \centering
  \includegraphics[width=\linewidth]{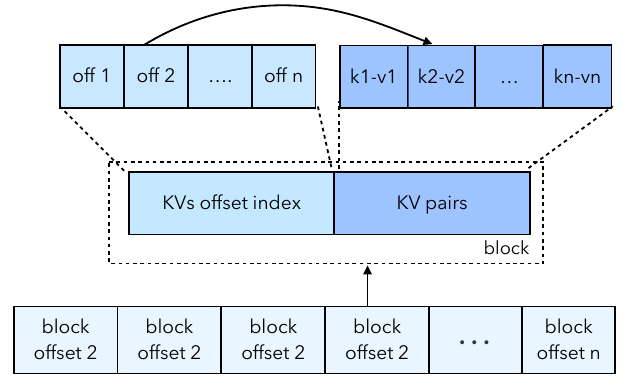}
  \caption{ The Block offset table and corresponding block. The block head stores a KV pair offset table which records the offsets of the KV pairs relative to the block.}
  \Description{h}
  \label{fig:blockoffset}
\end{figure}

To efficiently manage data and reduce index size, we organize keys into two levels of hierarchy, which are segments as described previous section and blocks. Each segment contains multiple blocks and each block contains fixed number of keys as shown in Figure\ref{fig:blockoffset}. Block is the minimal unit of each read operation. Typically, file systems read and write to the disk in units of pages (4KB usually). So we expected that the block size is close to page size to reduce I/O overhead.  We choose the fix number $k = \max(\lfloor\frac{page\;size}{avg(kv\;size)}\rfloor, 1)$, where average key-value size is obtained from previous statistics information.

After the $k$ is determined, we assemble every $k$ key-values into a block in slot index order.
All keys within one block share a block index entry which point to the block start offset in piece file, we only need keep block offset index entry in memory which can greatly save on memory usage.
When lookup a key, first to get the $slot$  by PH function, and get the block offset entry from $block\_offset\_table[\lfloor \frac{slot}{k} \rfloor]$ in memory, then reretrieve block from piece file where index point to, finally get actual key-value from where $kvs\_offset [slot\;\%\;k]$.

Organizing data into blocks also brings other benefits. Compression can be performed block by block, allowing for more compact storage. At the same time, data is written to the file in blocks to improve the speed of writing and mitigate disk fragmentation issues \cite{chen2009understanding,park2023filesystem}.

CompassDB’s meticulous design of piece file structure and block ensures that even when handling large-scale datasets, the database maintains exceptional lookup efficiency. With the index information residing in memory, CompassDB significantly reduces disk access frequency, optimizing data processing speed. This storage architecture is the fundamental reason CompassDB can provide fast and reliable data access performance.

\subsection{Index in Memory}
\label{ph_index_in_memory}
To quantify the minimal space occupied by the index in memory, we can measure the size of the index information that needs to be loaded into memory when searching for a key. In CompassDB, the following data from the global table needs to be loaded into memory:

\begin{itemize}
    \item The perfect hash parameters($\alpha$  $\beta$ pairs) average to about 1.2 bytes per key.
    \item Signature: Each signature occupies 8 bits, or 1 byte.
    \item Piece Index Table: Considering that the number of piece files is usually not very large (with a maximum limit of 1024), each key occupies about 1.25 bytes.
\end{itemize}
Combining the above information, the total number of bytes for the global hash table part is $(1.2 + 1 + 1.25) * 1.1 = 3.8 $bytes. Here, 1.1 represents the 0.1 redundancy slots in the actual perfect hash table.

Next, we calculate the local hash table part:

\begin{itemize}
    \item Similar to the global hash context, it averages 1.2 bytes per key.
    \item Each block contains 16 key-value pairs, with an average space consumption of 4 bytes / 16 = 0.25 byte per key-value pair.
\end{itemize}

Therefore, the total memory consumption for the local hash table part is $(1.2 + 0.25) * 1.7 * 1.1 = 2.71$ bytes. Here, 1.7 is the assumed space amplification, considering the proportion of keys being overwritten.
Combining the memory overhead of global and local hash tables, the memory consumption required to access a key is only $3.8 + 2.71 = 6.51$ bytes.

In CompassDB, each key is located based on its hash value, meaning that the memory occupied by metadata is independent of the actual size of the key-value pairs and only depends on the number of keys. This design allows CompassDB to maintain stable performance even as the data scale continues to grow. Actual tests have shown that when the value size is 300B and the dataset is 50 million, CompassDB’s memory usage is approximately 1.7G. Even when the dataset increases to 100 million, memory usage only reaches 2.9G (considering the existence of memtable, the calculation result may have some deviation).

\subsection{More Configuration}
\label{more configuration}

We recognize the diverse and complex nature of real-world business scenarios. Instead of pursuing a "one-size-fits-all" configuration that may introduce unnecessary overhead, CompassDB is dedicated to offering a range of optional configurations to cater to different business challenges and optimize performance for users.

\subsubsection{One Level}
For scenarios prioritizing high read performance with minimal write operations, CompassDB provides the optimized "one level" mode. This configuration streamlines the LSM-Tree structure by reducing it to just 1 or 2 layers. During memtable flush, Level-0 is bypassed, and multiple memtables are merged directly into Level-1. Utilizing CompassDB’s TPH delta piece file mechanism ensures that even when data overlaps with all TPHs in Level-1, write amplification remains extremely low. Data is segmented and integrated into existing TPH piece files in Level-1, approximately matching the size of the memtable being written. With a default upper limit of 16 piece files, CompassDB significantly reduces write amplification compared to RocksDB—one-sixteenth of the frequency, as unmodified keys in CompassDB are written only once versus multiple rewrites in RocksDB. While this mode requires reading all data during compaction, SSD read speeds, being faster than write speeds, make this trade-off feasible.

\begin{figure}[t]
  \centering
  \includegraphics[width=\linewidth]{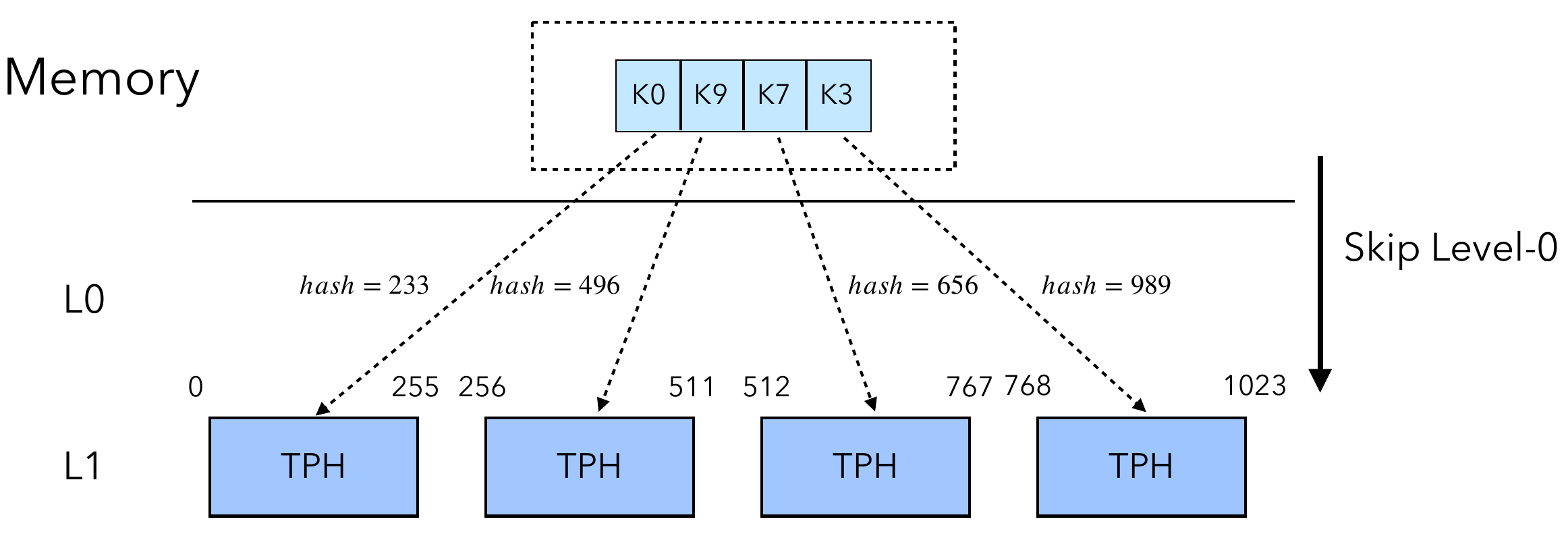}
  \caption{ The key-value pairs in memtables are directly merged into L1 during flush, L0 is skipped.}
  \Description{h}
  \label{fig:img-10}
\end{figure}

Accumulating multiple memtables before triggering compaction and utilizing page cache further mitigate read amplification. Tests demonstrate that handling 100 million keys with 1KB values shows RocksDB’s write amplification at 15 times that of CompassDB, with CompassDB reading about 40\% less data. Therefore, for read-intensive applications, CompassDB’s one level mode efficiently balances high-performance read operations with low write amplification.

\subsubsection{Single Tier}

In environments characterized by frequent overwrites, CompassDB offers the single tier mode. This mode addresses space and read amplification concerns by setting a smaller upper limit for delta pieces, such as 2 or even 1. This approach minimizes storage of outdated data, reducing unnecessary read amplification during access as the system seeks the latest data version across fewer piece files. Leveraging the perfect hash mechanism in TPH ensures efficient read performance. Tests indicate that under a dataset of 200 million key-value pairs with 1KB each, CompassDB achieves six times the read QPS compared to RocksDB in random read scenarios. The single tier mode optimizes storage space consumption while enhancing read performance, making it ideal for applications requiring rapid data updates and access.

These configurable options enable CompassDB to effectively address diverse business requirements and challenges in industrial-grade applications, ensuring optimal performance tailored to specific operational needs.

\section{Evaluation}
CompassDB is primarily written in C++ and serves mainly as an embedded KV storage engine. It is fully compatible with RocksDB, the most popular persistent key-value store.

In this section, we evaluated CompassDB read and write performance with two common benchmark tools for kv stores,  \verb|YCSB| \cite{cooper_brianfrankcooperycsb_2024} and \verb|db_bench| \cite{cao_characterizing_nodate}.

The evaluation is primarily based on three metrics:

\begin{itemize}
    \item   Throughput of operations.
    \item   Average latency and p99 latency for read and write operations.
    \item   Read and write amplification factors.
\end{itemize}

We compare CompassDB against RocksDB and PebblesDB. 
RocksDB, a leading high-performance persistent key-value store, is optimized for fast, low latency storage such as flash drives and high-speed disk drives, and adaptable to different workloads. PebblesDB is also a write-optimized key-value store which is built on Fragmented Log-Structured Merge Trees (FLSM) data structure. FLSM is a variant of the standard LSM-Tree data structure which aims at achieving higher write throughput and lower write amplification without compromising on read throughput, the same goal of CompassDB. In the experiment In this experiment, we use RocksDB version 7.9.2 and PebblesDB at commit \verb|703bd0|. 

\subsection{Setup}
We conducted experiments using a Dell PowerEdge T440 Tower Server.
Server equipped with an Intel Xeon Silver 2.40 GHz processor, a Samsung PM9A3 3.84TB NVMe SSD, and 128GB of RAM. The server runs Ubuntu 20.04 LTS with the EXT4 file system \cite{mathur2007new} in ordered mode.

To ensure the reliability of our experiment results and minimize the influence of external factors, we disabled the page cache, opting for direct IO (\verb|O_DIRECT| flag) for all disk reads \cite{open2_nodate}. This decision was made to accurately assess the impact of optimizations on disk performance, as caching data in memory could mask these effects. Additionally, compression was disabled during testing.

For stores configurations RocksDB and PebblesDB were configured with an LRU \verb|block_cache| set to 32MB each, whereas CompassDB currently operates without a block cache. All databases utilized up to 4 memtables, each capped at 128MB capacity, and maintained a 6-level LSM-Tree structure. CompassDB utilized the \verb|hash_range| mode to maximize query efficiency, while other configurations remained at their defaults.

\subsection{YCSB}

\begin{figure*}[t]
\centering
\subfigure[20 B key, 1KB value, 16 threads under different YCSB workloads.] {
\begin{minipage}[t]{0.5\linewidth}
\centering
\includegraphics[width=\linewidth]{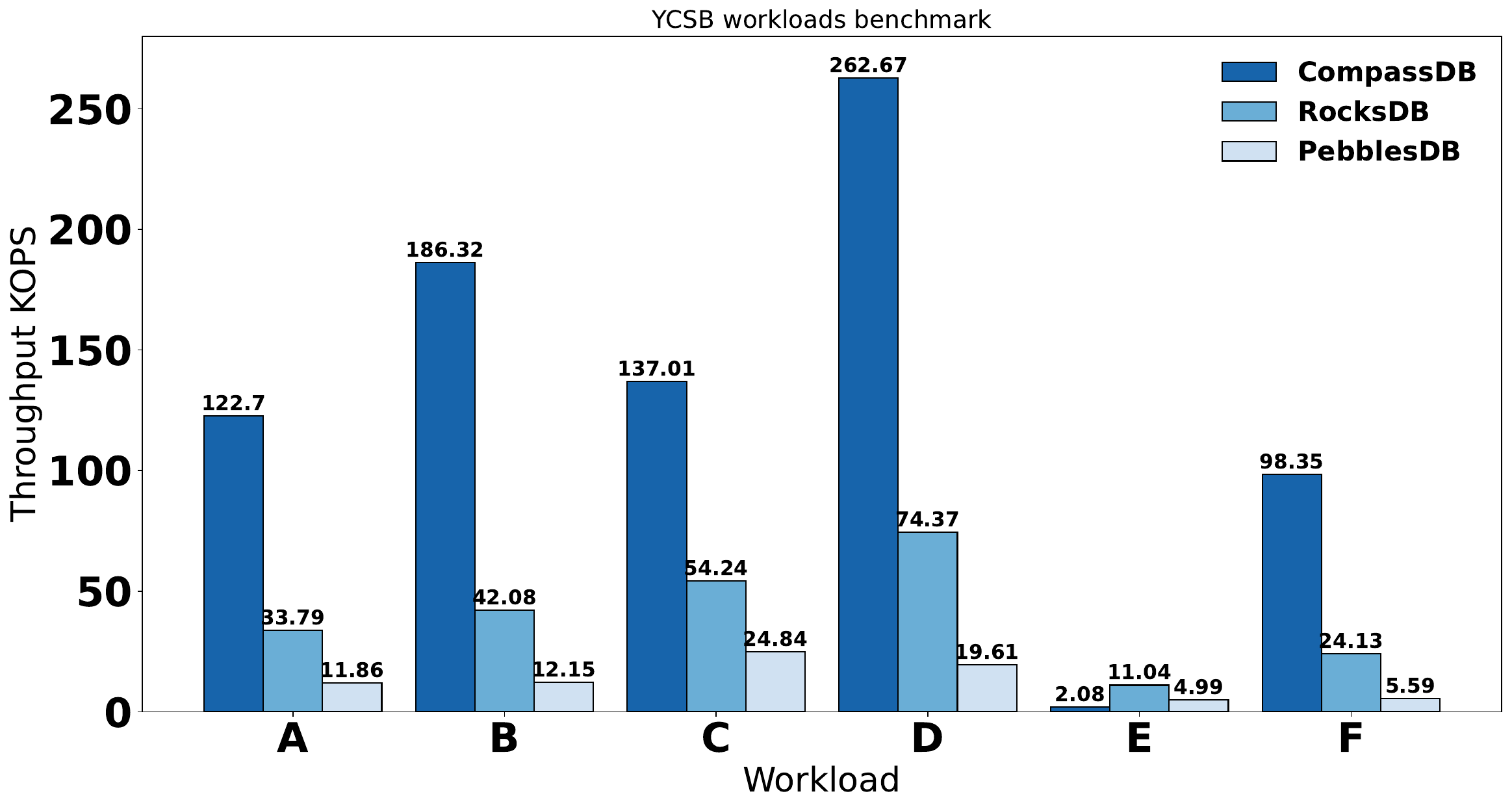}
\end{minipage}%
}%
\subfigure[20 B key, 16 threads, YCSB workload C under differen value size.]{
\begin{minipage}[t]{0.5\linewidth}
\centering
\includegraphics[width=\linewidth]{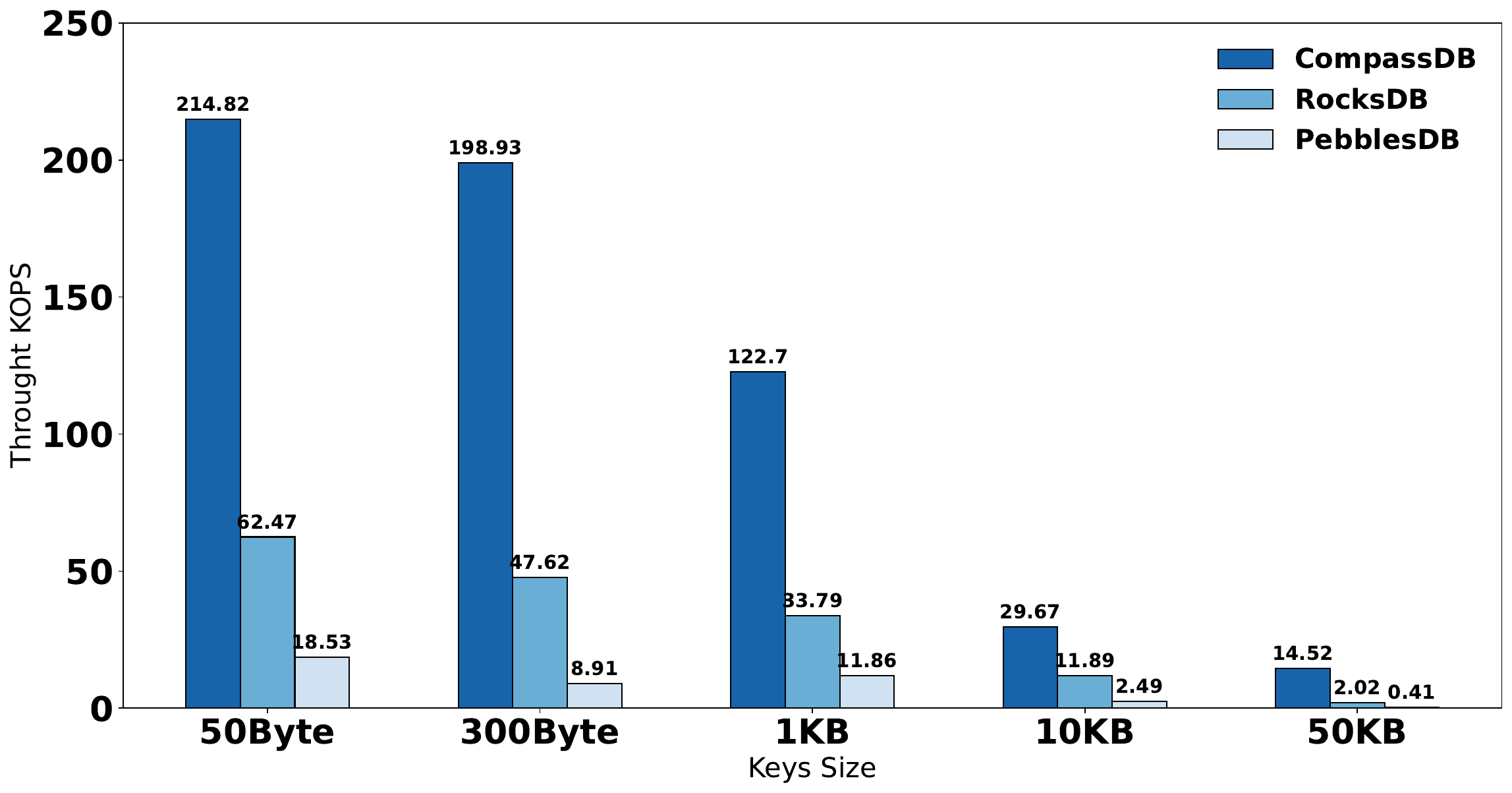}
\end{minipage}%
}%
\caption{CompassDB,RocksDB and PebblesDB throughput (KOPS) under different YCSB configuration. }
\Description{}
\label{fig:ycsb}
\end{figure*}
\begin{table}[t]
    \renewcommand\arraystretch{1.5}
    \begin{tabular}{cc}
     \toprule
    \multicolumn{1}{l}{Workload} & Description                           \\ \hline
     A& 50\% reads, 50\% writes
    \\ \hline

    B& 95\% reads, 5\% writes\\ \hline
    C                            & 100\%reads                            \\ \hline
    D                            & 95\% reads(latest values), 5\% writes \\ \hline
    E                            & 95\% Range queries, 5\% writes        \\ \hline
    F                            & 50\% reads, 50\% read-modify-writes   \\ 
    \bottomrule
    \end{tabular}
   \caption{A description of the YCSB workloads.}
   \label{table:ycsb-workloads}
\end{table}

    \begin{table}[t]

        \begin{tabular}{@{}clcccc@{}}
        \toprule
        \multirow{2}{*}{Workload} & \multirow{2}{*}{Engine} & \multicolumn{2}{c}{Read(us)}  & \multicolumn{2}{c}{Update(us)} \\ \cmidrule(l){3-6} 
        &                         & avg           & p99           & avg           & p99            \\ \midrule
    \multirow{3}{*}{A}        & compassdb               & \textbf{105}  & \textbf{862}  & \textbf{146}  & \textbf{1000}  \\
        & rocksdb                 & 442           & 1617          & 493           & 1673           \\
        & pebblesdb               & 1302          & 9279          & 1383          & 9503           \\ \midrule
    \multirow{3}{*}{B}        & compassdb               & \textbf{81}   & \textbf{289}  & \textbf{196}  & \textbf{348}   \\
        & rocksdb                 & 373           & 1623          & 432           & 1690           \\
        & pebblesdb               & 1297          & 18239         & 1401          & 18623          \\ \midrule
    \multirow{3}{*}{C}        & compassdb               & \textbf{113}  & \textbf{231}  & -             & -              \\
        & rocksdb                 & 289           & 613           & -             & -              \\
        & pebblesdb               & 637           & 18207         & -             & -              \\ \midrule
    \multirow{3}{*}{D}        & compassdb               & \textbf{59}   & \textbf{375}  & \textbf{35}   & \textbf{52}    \\
        & rocksdb                 & 219           & 1052          & 53            & 100            \\
        & pebblesdb               & 835           & 17935         & 110           & 346            \\ \midrule
    \multirow{3}{*}{E}        & compassdb               & 8027          & 15071         & \textbf{51}   & \textbf{98}    \\
        & rocksdb                 & \textbf{1513} & \textbf{2549} & 62            & 142            \\
        & pebblesdb               & 3355          & 8015          & 70            & 185            \\ \midrule
    \multirow{3}{*}{F}        & compassdb               & \textbf{97}   & \textbf{665}  & \textbf{120}  & \textbf{672}   \\
        & rocksdb                 & 423           & 1528          & 463           & 1568           \\
        & pebblesdb               & 3227          & 34495         & 752           & 16735          \\ \bottomrule
        \end{tabular}
\caption{CompassDB, RocksDB and PebblesDB average and p99 latency result from six YCSB workloads.}
\label{table:ycsb-benchmarks}   
\end{table}

The \verb|YCSB| is an open-source specification and program often used to compare the relative performance of NoSQL database management systems under different workloads
Table \ref{table:ycsb-workloads} describes the six typical  workloads (A-F) in the YCSB suite. 

We conducted workload benchmarks using 16 threads, initially loading 200 million keys (20 bytes per key, 1KB per value) into an empty database, followed by 100 million operations as detailed in Table 1.

Figure \ref{fig:ycsb}(a) shows throughput results (thousands of operations per second): CompassDB outperforms RocksDB and PebblesDB in the five of workloads(A, B, C, D, F),  but performs less effectively in scan-dominated workload E. Table \ref{table:ycsb-benchmarks}  provides average and p99 latency for read/write operations across all workloads.


For Workload A, which includes a balanced mix of read and write operations, CompassDB performs approximately 3 times faster than RocksDB and 10 times faster than PebblesDB. Both average and p99 latencies for reads and writes are significantly lower in CompassDB compared to the other databases. These results highlight the optimizations in CompassDB that enhance performance in mixed workloads. Additionally, the p99 latency in CompassDB is approximately 44\% lower than RocksDB and 90\% lower than PebblesDB, indicating more consistent performance during read and write operations.

Under read-only Workload C, CompassDB achieves 2.52 times and 5.5 times better performance than RocksDB and PebblesDB, respectively. Unlike Workload A, where RocksDB and PebblesDB show significantly reduced latencies, CompassDB maintains similar performance levels, indicating its resilience to write-intensive scenarios.

Workloads B and D are skewed towards read operations (95\% reads). In Workload D, which focuses on reading the latest values, CompassDB achieves speedups of 5.5 times and 13.3 times over RocksDB and PebblesDB, respectively, while also performing well in Workload B. This advantage is attributed to CompassDB's ability to  efficiently retrieve data from memtables and upper-level files (mostly Level 0) due to its TPH design, which minimizes disk I/O and filters out false positives effectively.

Workload E, primarily involving scan operations, shows CompassDB underperforming compared to RocksDB and PebblesDB. This is because CompassDB implements scans via random reads, whereas RocksDB and PebblesDB store data sequentially and benefit from prefetching into \verb|block_cache|, thereby reducing disk I/O.

  \begin{figure*}[t]
    \centering
    \subfigure[write amplification.]{
    \begin{minipage}[t]{0.5\linewidth}
    \centering
    \includegraphics[width=\linewidth]{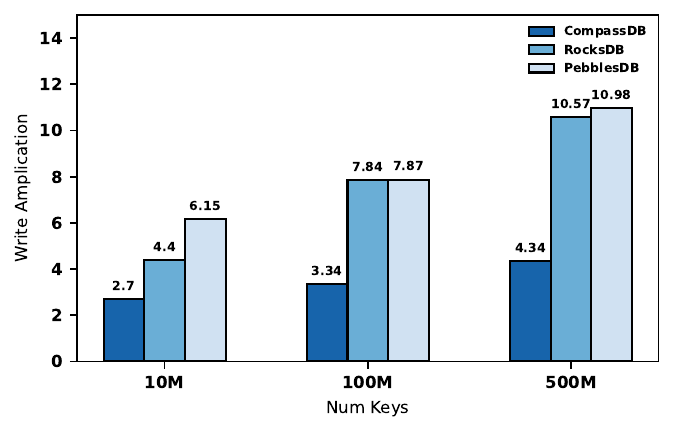}
    \end{minipage}%
    }%
    \subfigure[read amplification.]{
    \begin{minipage}[t]{0.5\linewidth}
    \centering
    \includegraphics[width=\linewidth]{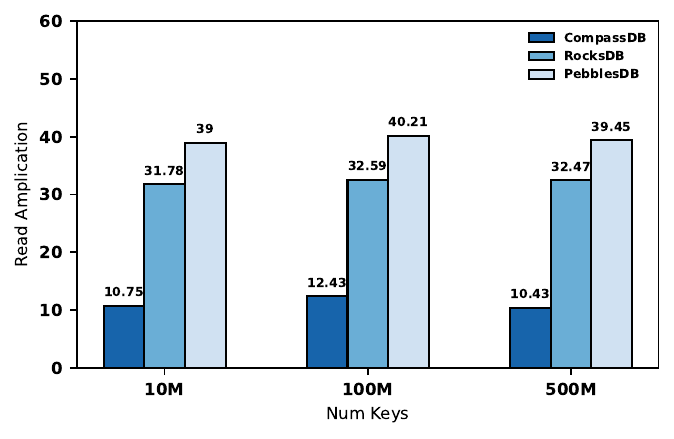}
    \end{minipage}%
    }%
    \caption{CompassDB,RocksDB and PebblesDB Write/Read amplicfication under different number of keys (20 B key and 128 B value). }
\Description{A woman and a girl in white dresses sit in an open car.}
 \label{fig:dbbench}
\end{figure*}

Workload F is similar to Workload A but involves update operations based on previously read data. All three databases perform similarly to Workload A with slight performance degradation.

Additionally, we tested Workload C with varying value sizes from 50 bytes to 50KB. Figure \ref{fig:ycsb}(b) depicts the throughput results. As value size increases, all databases experience decreased throughput. However, CompassDB shows increasingly better performance relative to RocksDB and PebblesDB. This advantage stems from the fact that while RocksDB and PebblesDB index blocks grow with larger values, requiring more time for searches, CompassDB's perfect hash index size depends only on the number of keys. Thus, CompassDB exhibits superior performance, especially with larger key-value pairs.

In summary, for point lookup and insertion workloads(A,B,C,D,F), CompassDB’s throughput is about 5 to 17 times faster than PebblesDB and about 2.5 to 4 times than RocksDB, and it also has more stable latency. With larger value sizes, the advantages of CompassDB over other databases become increasingly evident. For scan-dominated workloads, CompassDB do not perform as well as other DBs due to it has different access patterns. And this is part of our future optimization.


\subsection{micro benchmarks}

\verb|db_bench| is another commonly used benchmark tool for key-value stores, which can better simulate the real-world workloads for key-value stores. This benchmark is capable of synthesizing more precise key-value queries, which represent the read and write operations of key-value stores to the underlying storage system.

In this section, we utilize \verb|db_bench| micro benchmark tools to test the three databases, analyzing the differences in read and write amplification as well as read and write performance under various data volume sizes for each database.

\subsubsection{Write amplification} Write amplification refers to the ratio of the number of bytes written to storage compared to the number of bytes inserted into the database.
\begin{equation}
    wa =  \frac{bytes \: write \:to \: disk}{bytes \: write \: by \: user} 
\end{equation}

We use \verb|db_bench| to insert key-value pairs into three DBs in random order (\verb|fillrandom|). The size of the key is 20 bytes, and the value is 128 bytes. Thread number set to 1, it is not affect the result. 

Figure \ref{fig:dbbench} a reports the write amplification factor of the three DBs under different amount of keys. We observed that as the data volume increases, write amplification also increases. That is due to each level contains more data when amount of data grows, so when key-value pairs flows to the most bottom level, it maybe require more data to compact with. CompassDB keep the lowest write amplification across all of the worklads. As the amount of written data increases, the advantage of CompassDB’s write amplification is more pronounced compared with other DBs. For 500M keys, CompassDB lowers by 2.52x than RocksDB and PebblesDB. This demonstrates the effectiveness of the piece file structure which reduce the rewritten size of duplicated data.

  \begin{figure}[t]
    \centering
    \includegraphics[width=\linewidth]{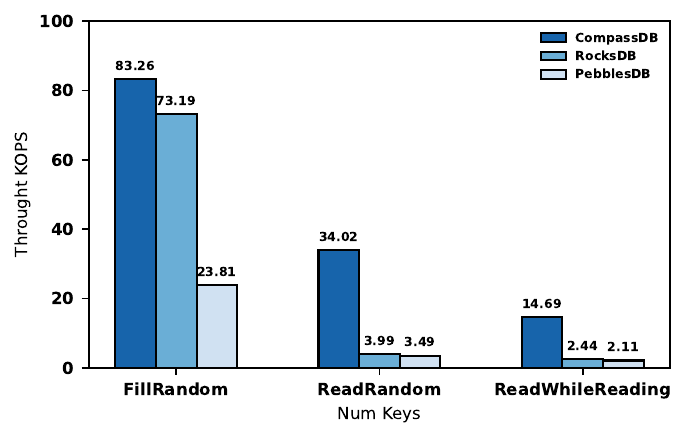}
    \caption{Throught (KOPS) under different workloads (FillRandom, ReadRandom and ReadWhileReading) with 200 million keys, 20 B key and 128 B value. }
    \Description{}
    \label{fig:db-bench-mix}
  \end{figure}

\subsubsection{Read amplification} Read amplification refers to the ratio of the actual amount of data read from the disk to the amount of data requested during a read operation. 
\begin{equation}
    ra =  \frac{bytes \: from \: disk}{bytes \: read \: by \: user} 
\end{equation}

A higher read amplification factor usually lead to poorer read performance. Under the data previously written, we evaluated the read amplification by data volumes by reading the entire key-value pairs of DB in random order by \verb|Get| interface.

 From the figure \ref{fig:dbbench} b,  it can be observed that the read amplification factor of CompassDB is close to 10, whereas RocksDB and PebblesDB are approximately 3x and 4x higher than CompassDB respectively. This is primarily because CompassDB requires no more than a single disk operation when reading from the PHTable. Additionally, PH index filter has lower false-pasitive rate ( $ 1/255 \approx 0.3\% $ ) compared to bloom filter ( $~1\%$ with 10bits/key), it can avoid most false-positive reads, thus preventing unnecessary disk operations.

\subsubsection{mix workload} 
We tested several common workloads using \verb|db_bench|, including \verb|fillrandom| for randomly inserting data, \verb|readrandom| for randomly reading data, and \verb|writewhilereading| for mixed read-write testing. 

The figure\ref{fig:db-bench-mix} show the throughput of the three databases during the tests. It can be observed that the insertion speed of CompassDB and RocksDB is similar during FillRandom, mainly because the database directories are empty during FillRandom, allowing data to flow from upper layers to lower layers at a faster rate.
At the same time, it is evident that CompassDB and RocksDB are over 3x faster than PebblesDB, which is largely due to the optimizations of CompassDB and RocksDB for concurrent writes. In the pure read workload, PebblesDB and RocksDB perform similarly, while the pure read efficiency of CompassDB is about 8 times that of the others. In the mixed read-write test, CompassDB is approximately 5.8 times faster than RocksDB and PebblesDB. These benefits are primarily due to CompassDB’s faster index filtering efficiency, fewer IO operations, and lower write and read amplification factors.

\subsection{CPH benchmark}

\begin{figure}[t]
    \centering
    \includegraphics[width=\linewidth]{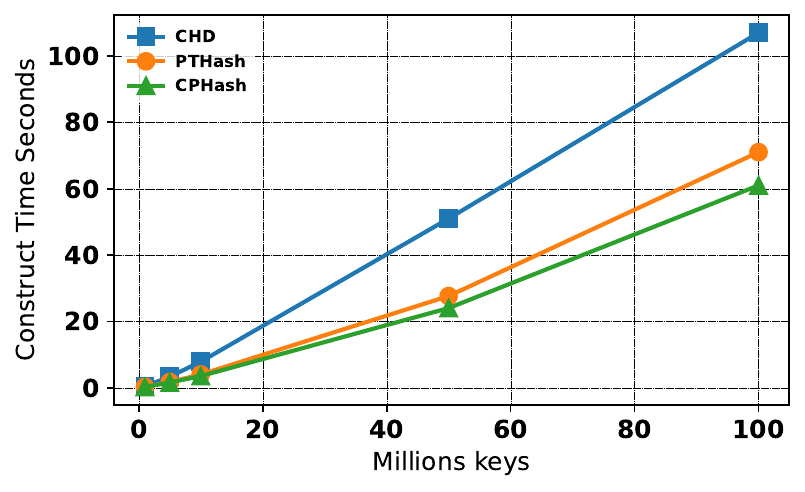}
    \caption{Performance of construction on different number of keys.}
    \Description{A woman and a girl in white dresses sit in an open car.}
  \end{figure}

CPHash is an algorithm that builds upon the CHD algorithm  by integrating the bucket dispersion concept from PTHash, and it employs optimization methods such as SIMD parallel computing and parameter sharing to enhance computational efficiency.

In our experiments, we set the load factor of the hash table to 0.9 and the average number of buckets (\verb|bucket_num|) to 5.  We evaluated the computational performance of these algorithms using a single thread and varying data volumes. As shown in Table 5.5-1, CPHash consistently demonstrated superior computational efficiency compared to the other two algorithms. Specifically, CPHash reduced time consumption by approximately 40\% compared to the CHD algorithm. Moreover, as data volumes increased, CPHash exhibited even lower time consumption relative to the PTHash algorithm. For instance, at a data volume of 100 million, CPHash reduced time consumption by approximately 14\%.

\section{Related Work}


\textbf{WiscKey} \cite{lu2017wisckey}  innovates by separating key and value storage, with values stored in a dedicated virtual log (vlog) replacing the traditional write-ahead log (WAL) in the Log-Structured Merge-Tree (LSM). Only keys and indices pointing to values are stored in the LSM, reducing unnecessary data writes during compaction and minimizing write amplification. This approach also reduces overall LSM volume, enhancing cache efficiency.

Derived from LevelDB, WiscKey requires multiple I/O accesses to metadata and one for the value during each read operation. Both CompassDB and WiscKey experience space amplification in scenarios with frequent key overwrites: WiscKey's vlog accumulates outdated entries, similar to how CompassDB manages piece files, necessitating garbage collection for storage release. Despite increasing data volumes, both systems maintain low read amplification; WiscKey achieves this with its smaller LSM, resulting in fewer levels and improved access performance.

CompassDB consistently offers $O(1)$ IO access complexity regardless of dataset size. Both systems require substantial random I/O operations during scans, necessitating asynchronous and parallel I/O techniques to boost performance. WiscKey excels in scenarios with small keys and large values, whereas CompassDB accommodates various key-value patterns effectively.

\textbf{PebblesDB} utilizes a novel data structure known as FLSM, managed through a mechanism called "guards." Each guard corresponds to an entire range and is logically composed of multiple SSTable files within that range. During compaction, the data of the guard for level $L_n$ is fragmented and organized before being directly added to the guard of level $L_{n+1}$ , thus reducing the amount of data rewriting. PebblesDB demonstrates good performance in terms of write amplification, write throughput, and read throughput. However, to reduce write amplification, it has made some compromises on read performance.

PebblesDB significantly reduces data rewriting by using guard. CompassDB adopted a piece-based approach to address this issue. When data from level $L_n$ is compacted to $L_{n+1}$, the delta data is directly turned into piece files for the next level. If the number of piece files exceeds a limit, a internal garbage collection is performed to release storage space. PebblesDB sacrifices read performance to reduce write amplification. CompassDB benefit from two-tier indexing structure, once the piece files are generated, the read performance remains unchanged, a read operation typically involves only a single read operation.


Because the guard of level $L_n$ must exist in $L_{n+1}$, which naturally divides the entire tree into multiple independent \verb|subtrees|. Each \verb|subtree| whose compactions do not affect each other, allowing for more tasks to be executed concurrently.CompassDB also has a similar mechanism. When using the hash range mode, it divides the data distribution according to the hash, with a fixed fan-out number file at each level. Therefore, when data from $L_n$ is compacted to $L_{n+1}$, the number of associated files is fixed, resulting in hight compaction concurrency.

\textbf{HashKV} \cite{chan_hashkv_nodate} is a solution proposed to address the inefficiency of WiscKey when dealing with update-intensive workloads. In WiscKey, where keys and values are stored separately, frequent update operations can lead to high write amplification and performance degradation. HashKV improves the efficiency of updates and garbage collection by employing a hashing method to deterministically locate the storage position of values.

In HashKV, each key is mapped to a specific storage location through a hash function, allowing for rapid access and updates to the corresponding value. This approach reduces the amount of data that needs to be read and rewritten during update operations, as the system does not have to search for and replace data throughout the entire data structure. Furthermore, the hash index also makes the garbage collection process more efficient, as the system can quickly identify which data is obsolete and needs to be purged.

\section{Conclusion}

In this paper, we present CompassDB, an industrial-grade key-value storage engine with high throughput and low latency. CompassDB primarily benefits from two designs: 1) CPHash index, a perfect hash function that provide $O(1)$ retrieval complexity. 2) TPH based on piece file design that reduce duplicate data rewritten to lower write amplification. Benchmarks show that CompassDB outperforms the leading and widely used key-value stores such as RocksDB and PebblesDB on several workloads. CompassDB is compatible with all RocksDB API, users can easily migrate their applications to CompassDB.



\bibliographystyle{ACM-Reference-Format}
\bibliography{compassdb}

\end{document}